\documentclass[amscls]{revtex4}
\newcommand{\be}{\begin{equation}}
\newcommand{\ee}{\end{equation}}
\newcommand{\ba}{\begin{eqnarray}}
\newcommand{\ea}{\end{eqnarray}}
\begin{document}
%\draft
%%%%%%%%%%%%%%%%%%%%%%%%%%%%%

\title{Near-Horizon Geometry and the Entropy of a Minimally
Coupled Scalar Field in the Schwarzschild Black Hole}

\author{Kaushik Ghosh\footnote{E-mail ghosh{\_}kaushik06@yahoo.co.in}}
\affiliation{Vivekananda College,University of Calcutta, 
269, Diamond Harbour Road, Kolkata - 700063, India}

\maketitle

\section*{Abstract}
%\vspace{2cm}

In this article, we will discuss a Lorentzian sector calculation of the entropy 
of a minimally coupled scalar field in the
Schwarzschild black hole background using the brick wall model of t' Hooft.
In the original article, the Wentzel-Kramers-Brillouin (WKB) approximation was used for the
modes that are globally stationary. In a previous article, we found
that the WKB quantization rule together with a proper counting of the states,
leads to a new expression of the scalar field entropy which is not
proportional to the area of the horizon. The expression of the entropy
is logarithmically divergent in the brick wall cut-off parameter in contrast 
to an inverse power divergence obtained earlier. 
In this article, we will consider the entropy for a thin shell of matter field of 
a given thickness surrounding the black hole horizon.
The thickness is chosen to be large compared with the Planck length
and is of the order of the atomic scale.
We will discuss the corresponding boundary conditions and the appropriateness
of the WKB approximation using the Regge\textendash Wheeler tortoise coordinates.
When expressed in terms of a covariant cut-off parameter,
the entropy of a thin shell of matter field of a given thickness and surrounding the horizon
in the Schwarzschild black hole background is given by an expression proportional to the area of the black hole horizon.
This leading order divergent term in the cut-off parameter remains to
be logarithmically divergent. The logarithmic divergence is expected from the nature
of the near-horizon geometry and is discussed in detail at the end of the section:II. 
We will find that these discussions are significant in the context of the continuation to the Euclidean sector
and the corresponding regularization schemes used to evaluate the thermodynamical properties of matter
fields in curved spaces. These are related with to geometric aspects of curved spaces. The above discussions are also
significant in presence of cosmological event horizon.

Key-words: black holes and scalar field entropy, brick wall model, area law, logarithmic divergence, 
$\zeta$-function regularization.

PACS numbers: 97.60.Lf, 04.70.Dy, 04.62.+v, 11.10.Wx 	

%MSC numbers: 81Q20, 81T20, 58J35, 11M36.

%\end{Abstract}

\newpage

\section {I\MakeLowercase{ntroduction}}

The thermodynamical aspects of the black holes was first established by
Bekenstein [1]. He obtained an expression for the entropy of the black
holes. This expression was proportional to the area of the horizons
of black holes. Hawking obtained the exact expression of the
entropy by considering the behaviour of matter fields in the black
hole background [2]. The entropy of a black hole, considered as a
thermodynamical system, was found to be ${A\over{4}}$. Here,
$A$ is the horizon surface area. The thermodynamical properties of a
minimally coupled matter field in the black hole backgrounds were discussed by 't Hooft [3]. 
The black hole is assumed to be in thermal equilibrium with the
surrounding matter. For a static black hole, the rate of particles
radiated by the black hole is equal to the rate of the absorption
of matter by the black hole. 't Hooft [3] assumed that the matter
field wave function is vanishing near the horizon as well as at
infinity, i.e. for a large value of the radial coordinate. Since
then, a lot of works had been carried out on the entropy of matter fields in
the black hole backgrounds using the brick wall model [4 \textendash 13].

In a previous article [14], we discussed a few aspects of the entropy
of a scalar field in the Schwarzschild black hole
background using the brick wall model of 't Hooft [3] with a proper counting of the states.
We discuss these calculations briefly in Section:II.
In the brick wall model of 't Hooft, the semiclassical WKB quantization rule is applied to
the radial part of the matter field solution for the purpose of 
counting the states. The wavenumber depends on the radial variable. This
radial dependence of the wavenumber is associated with the `centrifugal potential' and
the metric. The `centrifugal potential' is vanishing at the horizon as well as at infinity
with respect to an asymptotic observer at infinity. 
There is a maximum at an intermediate value of the radial coordinate. The above terminology is used following Eqn.(15).
We considered the facts that the modes are the eigenstates of the angular momentum, and, if
the WKB quantization rule used in the brick wall model [3] remains to be valid, they
should be stationary throughout the range of the radial variable under consideration
for each angular quantum number.
This determines the allowed values of the angular quantum number.
We found that this consideration led to an expression of the entropy different from that obtained in [3].
The expression of entropy obtained by 't Hooft is inversely divergent in the brick wall
cut-off parameter and, in terms of a proper distance cut-off parameter, is
proportional to the area of the event horizon. The new expression of the
scalar field entropy obtained for these modes in [14] is logarithmically divergent
in the brick wall cut-off parameter and is not proportional to the area
of the black hole event horizon. We also note that the term logarithmically
divergent in the brick wall cut-off parameter is independent of the mass
of the black hole if we choose the temperature to be given by
the Hawking temperature.

The free energy of the globally stationary modes in the Schwarzschild 
black hole background is given by Eqn.(9).
The leading-order divergent part in terms of the large distance
cut-off parameter is proportional to the large distance cut-off
parameter and is not related to the corresponding volume.
Similar behaviour is also observed in the flat space.
We also find that the free energy given by Eqn.(9) vanishes
in the limit of the flat spacetime $(M \rightarrow 0)$. These may be due to the WKB approximation
with $k(r)$ given by Eqn.(4) and also to the global
stationary nature of the solutions.
The form of $k(r)$ indicates that the WKB approximation
may hold well for the solutions that are not stationary throughout the 
spatial manifold. 
In the flat spacetime the entropy of a thin shell of scalar field, 
obtained using the WKB approximation used in this article, is proportional to the volume 
of the thin shell in the limit that the thickness is small compared with the radius
of the inner surface. The internal energy and entropy vanish when the
thickness vanishes. Thus, the WKB approximation holds well for a thin shell of
matter field in the flat space. It is well known that the gravitational entropy can be associated with
the fixed point sets of the timelike killing field [4,15,16]. In section:III, we
will consider the entropy of a thin shell of massless scalar field with a definite thickness
surrounding the black hole event horizon [3]. The thin shell is important
to illustrate the significance of the near-horizon geometry to the black hole thermodynamics.
We will find that the WKB approximation is particularly suitable in this case
and illustrate the corresponding boundary conditions
using the Regge\textendash Wheeler tortoise coordinates. This may also be envisaged
from the nature of the `effective potential' in the Schwarzschild black hole
expressed in terms of the Regge\textendash Wheeler tortoise coordinates as given
by Eqs.(14) and (15). We will find the entropy
of a thin shell of a massless scalar field of a given thickness
surrounding the horizon following the procedure discussed in section:II. 
The temperature is given by the Hawking temperature.
To the leading order in terms of a proper distance cut-off parameter,
the scalar field entropy for a thin shell surrounding the black hole horizon
is proportional to the area of the black hole event horizon and is given by the
same expression for both the Schwarzschild and the nonextreme Reissner\textendash Nordstrom black holes.
We will again find that for a thin shell with a given thickness,  
the free energy and entropy are logarithmically
dependent on the coordinate and the proper distance cut-off parameters. 
This is associated with the form of the solutions in the near-horizon region given by Eqn.(12) to Eqn.(15),
the WKB quantization rule with $k(r)$ given by Eqn.(4) or Eqn.(19), the
infinite redshift near the horizon and the form of the proper distance in the near-horizon region. 
This is discussed in detail at the end of the section:II.
Similar situation will remain valid with the nonextreme Reissner\textendash Nordstrom black holes.
It is remarkable that the entropy is given by a universal
expression proportional to the area of the horizon. 
The universality of the thin shell matter
field entropy with a given thickness leads to a universal expression for the covariant cut-off
parameter if we equate the matter field entropy to the black hole
entropy itself [3]. The covariant cut-off parameter is found to be extremely
small compared with the radius of the horizon of the black hole 
and is dependent on the thickness of the thin shell. If we choose
the thickness of the thin shell to be large compared with the Planck length
and of the order of the atomic scale,
the covariant cut-off becomes extremely small, {\it i.e.} the brick wall cut-off is
almost coincident with the horizon. This is expected if we consider the
solutions of the wave equation given in Section:III and is consistent
with the observed behaviours of the solutions in the optical metric
approach. In both cases it is found that there exist solutions
that are vanishing at the horizon [17,18,19]. 
The internal energy of the scalar field is less than
the mass of the black hole and is given by Eqn.(30). This behaviour is
different from those discussed in [3] where we can not push the brick wall
arbitrarily close to the horizon owing to the divergence of the scalar field internal energy.   
In the thin shell
model we have considered the solutions that are vanishing at the two boundaries
and are stationary throughout in between. Instead of using the rigid wall boundary conditions,
we can consider a half-infinite potential well and apply the WKB approximations.
The method of calculation presented in this article makes the
thin shell model expression a good approximation of this more realistic situation. 
We will find that the gross
features of the entropy as had been obtained with the thin shell model, namely, that it is proportional
to the horizon area and that it is logarithmically divergent in the proper distance brick wall
cut-off parameter remain unchanged with these half-infinite potential well boundary conditions.
Thus, the black hole entropy may be interpreted as originating from a
cluster of matter field confined in the near-horizon region with a thickness
large compared with the Planck length but small compared with the radius of the horizon.
This is important in the context of
explaining the black hole entropy in terms of the near-horizon states [4,16,17,20,21,22] and
also for the entanglement entropy approach to explain the black hole entropy [11,23].

The thermal behaviours of matter fields in different curved spaces have been
extensivly studied. One can look at the references [7,18,24] for a review.
The brick wall model is a Lorentzian sector calculation of the entropy of a matter
field in a black hole background. There are other approaches using the Euclidean
sector of the manifold.
Among these, the Euclidean path integral formulation [7,12,17,25,26,27] and the optical
metric approach [7,28] are noteworthy. The boundary conditions on the matter field
are similar to those used in the present article. In these approaches, the 
thermal properties of a scalar field in the near-horizon region are 
described in terms of the near-horizon geometry. We will give a brief comparative discussion
of these different approaches to evaluate the matter field entropy in a curved
space at the end of Section:III. We will also discuss the different interpretations of the matter field entropy 
at the end of Section:III.

\section {B\MakeLowercase{rick} W\MakeLowercase{all} M\MakeLowercase{odel and the} E\MakeLowercase{ntropy of a} 
S\MakeLowercase{calar} F\MakeLowercase{ield in the}
B\MakeLowercase{lack} H\MakeLowercase{ole} B\MakeLowercase{ackgrounds}}

The radial part of the wave equation of a massive scalar field in
the Schwarzschild black hole background is given by the following
expression:

\be
{(1 - {2M \over {r}})^{-1}}{E^2}{\psi(r)} + {1 \over {r^2}}
{{\partial}_r} [{r(r - 2M)}{{\partial}_r{\psi(r)}}] - [{{l(l +
1)}\over{r^2}} + {m^2}]{\psi(r)} = 0 .
\ee

\noindent {Here, $l$ is the angular momentum quantum number and $E$ is the
energy.}

We follow the approach of t' Hooft to calculate the entropy of a
scalar field using the brick wall model and the WKB quantization
rule [3]. We consider massless fields but the discussions can also
be extended to massive fields. The boundary conditions on the
scalar field are the following:

\be
{\psi(2M + h)}, {\psi(L)} = 0.
\ee

\noindent{Here, $h << 2M$ and $L >> 2M$. Here, $h$ is the brick wall cut-off
parameter. We consider the solutions of the wave equation obtained
by the WKB approximation. The trial solutions are assumed to be of the
form ${\rho(r)}{{\exp}[{i S(r)}]}$. The actual solutions are sinusoidal functions
to satisfy the boundary conditions.
We firstly consider the solutions that are stationary with respect to
the radial variable throughout the spatial manifold and the amplitude is
assumed to be a slowly varying function of the radial coordinate $r$.
These solutions are also important for the analysis of the Hawking radiation.}

We now consider the free energy calculation using a proper density
of states. Let us first consider the WKB quantization condition
together with the brick wall boundary condition. We have the
following condition:

\be
\pi n = {{\int_{r_1}}^{L}}{dr}k(r,l,E).
\ee

\noindent{Here, ${r_1} = 2M + h$ and $k(r,l,E)$ is the radial wavenumber.
Every state is characterized by a definite value of each of the three
quantum numbers $(n,l,m)$. The upper value of $l$ is
restricted by the condition that, for every choice of $l$, $k(r)$ should remain real
throughout the complete region of integration. We have}

\be
k(r) = {1\over{V(r)}}{\sqrt{{E^2} - {V(r)\over{r^2}}l(l + 1)}}.
\ee

\noindent{Here, we are considering a massless scalar field and $V(r) = (1 - {2M \over r})$,
where $M$ is the mass of the black hole. At ${r_1} = 2M + h$
with $h \rightarrow 0$, $l$ can be varied freely. However, for a
state for which $k(r,E,l)$ is real throughout the complete region
from $r = 2M + h$ to $r = L$, the upper limit of $l$ is restricted by
the maximum value of ${V(r)\over{r^2}}$, which occurs at $r = 3M$
for the Schwarzschild black hole. The upper limit of $l$ is given by the following expression:}

\be
J(J + 1) = s(27{{E^2}{M^2}}).
\ee

\noindent{Here, $s \leq 1$ so that $k(r)$ is real for \textit{each value of $l$} and for all values of $r$ in accordance
with the WKB quantization rule. We can estimate the value of $s$ from the usual condition for the validity
of the WKB approximation: ${|{\nabla[S(r)]}|^2} >> |{\nabla}^2 [S(r)]|$.
In this article, we are mostly interested in the near-horizon region. It can be
shown that for the macroscopic black holes, the WKB approximation holds well with $s << 1$. 
In this section, we will find that the higher order terms in $s$
are not much significant.}

The total number of solutions with energy not exceeding $E$ is then given by

\be
\pi N = g(E) = {{\int_{r_1}}^{L}}{dr}{\int (2l + 1){dl}}k(r,l,E).
\ee

\noindent{The expression of the free energy of a massless scalar field is
given by the following expression [3,29]:}

\be
{\pi}{\beta}F = -{{\int_{0}^{\infty}}{dE}{{\beta g(E)}\over {{\exp({\beta}E) - 1}}}}.
\ee

\noindent{In this case, we have}

\ba
{\pi}{\beta}F & = & - {{{\int}_{0}^{\infty}}{{{\beta}dE}\over{[{\exp}({\beta E}) -
1]}}} \\ \nonumber
& ~ & {{{\int}_{2M + h}^{L}}{dr\over{V(r)}}} {\int dl(2l + 1){\sqrt{E^2 -
{V(r)\over{r^2}}[l(l + 1)]}}}.
\ea

\noindent{The range of $E$ integration is from zero to infinity, although one should be considering the back
reaction to the metric. The range of $r$ integration is from $2M +
h$ to $L$, where $L >> 2M$. The $l$ integration in Eqn.(8) for the free energy 
introduces a factor ${{r^2}\over{V(r)}}$ in the radial integral, and we have}

\vspace{0.3 cm}

~~~~~~~~~~~~~~~~~~~~~~~~~${\pi}{\beta}F = - {2\over{3}}{\beta}{\int{{dE}\over{[{\exp}({\beta E}) -
1]}}} {\int{{{r^2}dr}\over{{V^2}(r)}}} {[E^3 - {\{{E^2 -
s(27{{E^2}{M^2}}){V(r)\over{r^2}}}\}}^{3\over{2}}]}$.

\vspace{0.3 cm}

\noindent{We can perform a binomial expansion in $s$ to determine the leading
divergent term in the brick wall cut-off parameter. The maximum
value of the term ${27{M^2}}{V(r)\over{r^2}}$ is one. We thereafter
do the radial integration and then the energy integration.
It can easily be seen that the term associated with the linear order-term
in $s$ is the most divergent in the brick wall cut-off
parameter $h$. The free energy is given by the following expression:}

\be
{\pi}F = -({{sI}\over {{\beta}^4}})(27{M^2})[L + 2M{\ln({L\over{h}})}],
\ee

\noindent{where $I = {{{\int}_{0}}^{\infty}}{{{x^3}dx}\over{({\exp{x}} -
1)}} = {{{\pi}^4}\over{15}}$ and $L >> 2M$. Note that the leading-order
infrared divergent term is not proportional to $L^3$. This is a shortcoming
of the WKB approximation. The free energy vanishes in
the flat spacetime $M \rightarrow 0$ limit. This is related to the WKB approximation
with $k(r)$ given by Eqn.(4). In the flat spacetime limit, the brick wall is
pushed at the origin. With $k(r)$ given by Eqn.(4), the number of the allowed
values of $l$ is of the order of $h^2$ and is vanishing in the approximation
considered here.}

The entropy is given by $S = {{\beta}^2}{{dF}\over{d
\beta}}$. The entropy associated with the linear-order term in $s$
is given by

\be S = {{27sI}\over{{128}{\pi}^4}}[{L\over{M}} +
2{\ln({L\over{h}})}]. \ee

\noindent{Here $\beta$ is given by the usual
expression $\beta = 8 \pi M$ and $h$ is the brick wall cut-off
parameter.
The divergent part of the entropy associated with the brick wall cut-off parameter
is logarithmically divergent in the
cut-off parameter. This is different from the corresponding first-order divergent
expression obtained by 't Hooft as given by Eqn.(23).
The divergent behaviour is expected as the constant-time foliations
intersect at the horizon and can be related with infinite redshift.
We have assumed $h << 2M$ and $L >> 2M$. We find that, with the above choice
of $\beta$, the divergent part of the entropy in terms of the brick wall cut-off parameter
is independent of the horizon surface area.
We also note that the term logarithmically
divergent in the brick wall cut-off parameter is independent of the mass
of the black hole, although the leading-order infrared divergent term is dependent
on the mass of the black hole. The higher order terms in
$s$ are not divergent in $h$ and vanish in the limit $L \rightarrow
\infty$.}

Note that, in the expression for $g(E)$, i.e., the total number of solutions with
energy less than or equal to $E$, we have multiplied $n$ with the angular degeneracy
factor whose upper limit is determined in terms of $E$. This corresponds to taking
the same angular degeneracy factor for all values of the radial quantum number below
$n$. In reality, the different values of $n$ have different allowed values for the 
maximum energy and the angular degeneracy factor should be different as is apparent from
Eqn.(5). This amounts to changing the value of the energy integral from the actual
value. However, when we choose an appropriate upper limit for the maximum
value of $l(l + 1)$ for a given value of $n$ (given by Eqn. 5), we find that after a
rescaling of the energy by the factor $\beta$, the radial quantum number $n$ scales like $\beta$ and the maximum allowed
value of $l(l + 1)$ scales like ${\beta}^2$. The free-energy integral will scale like
${\beta}^4$ since the lower limit of the integral is zero and the upper limit is infinity.
We will justify these limits in Section:III. Thus, when we set ${\beta}$ to be the inverse Hawking temperature, 
the form of the dependence of the scalar field entropy on the parameters of the black hole
will remain the same. A more accurate procedure to calculate the entropy will be discussed in section:III.

It is noteworthy that the leading infrared divergent part in the above
expression (9) is not proportional to $L^3$ as will be the situation in the
flat spacetime and in the Schwarzschild space for $L >> 2M$.
Similar situation remains valid with the flat spacetime.
If we calculate the entropy of a scalar field confined between
$r = R$ and $r = L$ with $L >> R$, the leading-order term is proportional
to ${R^2}L$, which is the volume of a box. This expression is only appropriate
for a thin shell of matter scalar field and we will discuss this issue further in Section:IV. 
This is a limitation of the WKB approximation and the semiclassical
quantization rule when $k(r)$ is expressed by Eqn.(4). In the flat spacetime, the centrifugal potential term
does not appear in the stationary phase parts of the exact solutions of the
nonrelativistic free-particle
wave equation [App.A, 30]. If we consider the field to be confined within
an annular region of finite width, the range of the allowed values of
$l$ with a given value of $E$ will be different. In this context, we note the
following. The nonrelativistic solutions in the flat space are expressed in terms of the spherical Bessel
and spherical Hankel functions. If we consider the matter field to be confined
within an annular region with arbitrary values of the two radii, it is not obvious 
to satisfy the boundary conditions when the solutions are expressed in terms of
linear combinations of the above mentioned special functions and are the
eigenstates of the angular momentum.

In passing, we make a few comments regarding the free energy calculation in the original 
article [3]. Equation (3) gives a single value of the radial quantum number
$n$ with a chosen range of the radial variable in which $k(r)$ is real, a definite value
of $E$ and a definite value of $l$. In [3], $k(r)$ is taken to be zero in Eqn.(3) when it
is not real. However, the upper limit of the $l$-integration in Eqn. (8)
was taken so that $k(r)$ is real up to $r$. This makes the upper limit of
$l$-integration dependent on the running variable $r$. We should ensure that we choose only those values of $l$
such that $k(r)$ is real throughout a chosen range of the radial variable 
for every single choice of $l$. Let us consider the case when $k(r)$ is real for the regions
$[{r_1},{R_1}]$ and $[{R_2},L]$, where ${{V(R_1)} \over {{R_1}^2}} = {{V(R_2)} \over {{R_2}^2}}$
(this case is mentioned briefly later in this section and in detail in [31] and does not lead
to interesting results). Here $R_1$ is less than $3M$. To calculate $g(E)$ or ${\pi}N$,i.e.,
the total number of solutions below energy $E$, we determine the
radial quantum number $n$ by the WKB approximation, 
multiply $n$ by $(2l + 1)$, and integrate over $l$
for all those values for which $k(r)$ is real in the 
chosen regions:

\be
{\pi}N(2M + h,L) = [{\int}{dl}(2l + 1)][{\int_{r_1}^{R_1}}{dr}k(r,l,E)
+ {\int_{R_2}^{L}}{dr}k(r,l,E)].
\ee

\noindent{We need not consider the region ${R_1} < r <{R_2}$ where $k(r)$ is
not real and we can set $k(r) = 0$ in the WKB quantization rule for this range of values 
of the radial variable $r$. In this case, the upper limit of the allowed values of $l$ is  
$J(J + 1) = {{{E^2}{R_1}^2} \over V(R_1)} = {{{E^2}{R_2}^2} \over V(R_2)}$.
Lastly, we put the angular integration in the radial integral. 
A running variable $r$-dependent constraint imposed on $l$ does not ensure 
the reality of $k(r)$ throughout the chosen range for the radial coordinate.
If we now want to extend the radial integral for all values of
$r$ from $r_{1}$ to $L$ to get Eqs.(6,8), the $l$ integration 
should be over only those values for which $k(r)$ is real for the complete 
range of $r$ from $r_{1}$ to $L$ and for every choice of $l$.
Again an $r$-dependent variable constraint imposed on $l$ does not ensure this.}

The ultraviolet divergent part associated with the brick wall cut-off of 
the scalar field entropy as obtained in [3] 
is given by the expression
$S = {{8{\pi}^3} \over {45 h}}{2M}({{2M} \over {\beta}})^3$.
Unlike Eqn.(11), this expression is inversely dependent on $h$.
In terms of a coordinate invariant cut-off, given by Eqn.(25),
this expression is proportional to the area of the horizon and is
quadratically divergent in the proper distance brick wall cut-off
parameter. The expression is independent of the mass of a particular black hole.
In section:IV of the present article, we
find that the approach followed by us gives more appropriate results. To explain this,
let us consider the solution in the near-horizon region 
discussed in Section:IV of the present manuscript. It is obvious from
Eqn.(4) or from Eqs.(13),(14) and (15) obtained by using the Regge\textendash Wheeler
tortoise coordinates and is a better approach to study the wave equation
in the near-horizon region that, in the near-horizon region, we can set
$V(r) = 0$ in the numerators as a first approximation. We can now use the WKB approximation
to a thin shell of scalar field confined in the near-horizon region.
It is obvious that the $l$-integration is free and $g(E)$ is only
logarithmically divergent in the brick wall cut-off parameter. Surely,
a constraint on $l$, when properly imposed, can not make $g(E)$ more
divergent in terms of the brick wall cut-off parameter. If we
look at Eqn.(27) of the present article we find that this aspect
is nicely incorporated through the presence of a ${1 \over {{\epsilon_d}^2}}$
factor, which indicates that, when the thickness of the thin shell is very 
small, the entropy is quadratically divergent in ${\epsilon_d}$,
the proper thickness of the thin shell. A quadratic divergence is expected 
from the upper limit when the $l$ integration is free. This is now
associated with the thickness of the thin shell and has nothing to
do with the brick wall cut-off parameter - a reasonable result indeed
from the nature of the solutions in the near-horizon region.

To conclude this section, we note that we can calculate the entropy associated with the modes
in the Schwarzschild black hole, which are stationary up to a certain value of $r = R_1$ $({R_1} < 3M)$
and again from $r = R_2$ $({R_2} > 3M)$ to infinity, where $R_1$ and $R_2$ are related by the
expression: ${{V(R_1)} \over {{R_1}^2}} = {{V(R_2)} \over {{R_2}^2}}$.
The solutions are not stationary in the intermediate region. We 
use the WKB quantization condition, Eqn.(3), to find the states with the upper
limit $L$ replaced by ${R_1}$. The expression ${{V(r)} \over {{r}^2}}$ is 
monotonically increasing in the range $r = 2M + h$ to $r = 3M$.
Hence, the upper limit of the $l$-integration is now is given by the
expression: $J(J + 1) = {{{E^2}{R_1}^2} \over V(R_1)} = {{{E^2}{R_2}^2} \over V(R_2)}$.
The upper limit of the
$l$-integration is divergent if we consider $R_1$ to be close to the horizon.
In the next section we will consider this issue in detail with a thin shell model.

The free energy and entropy can be calculated in the usual way [31] and to the
leading order terms, we have the following expression of the entropy:

$S = {{4{\pi}^3 s} \over {15{\beta}^3}}
[{{2M{{R_1}^2}} \over V(R_1)}{\ln({{R_1 - 2M}\over{h}})} + {R_1}^3]$.

\noindent{Note that the divergent term associated with the brick wall cut-off parameter
and the horizon is not much different from the previous expression of entropy obtained
for the globally stationary modes, eqn (10). We also find that, in the flat space limit 
$M \rightarrow 0$, the above expression vanishes
since we also have ${R_1} \rightarrow 0$ in this limit and this region no longer exists.
If we choose the temperature in the above expression to be given by the Hawking temperature,
$\beta = 8{\pi}M$, the logarithmically divergent term is not proportional to
the area of the horizon.}

The thermal behaviours of matter fields in different curved spaces have been
extensivly studied. One can look at the references [7,17] for a review.
There are a number of different approaches to find the entropy of a scalar field
in black hole backgrounds apart from the brick wall model of 't Hooft.
Among these, the optical metric approach and the Eucledian path
integral formulations are noteworthy [7,12,17,25,26]. In these approaches the 
thermal properties of a conformally coupled scalar field are 
described in terms of the near-horizon geometry. The near-horizon geometry is also significant in the
quantum gravitational approaches explaining the black hole entropy [4,16,17,20,21,22].
Thus, in Section:III, we will consider the entropy of a thin shell of matter field in the
near-horizon region surrounding itself.

\section {E\MakeLowercase{ntropy of a} T\MakeLowercase{hin} S\MakeLowercase{hell of} 
S\MakeLowercase{calar} F\MakeLowercase{ield} S\MakeLowercase{urrounding
the} B\MakeLowercase{lack} H\MakeLowercase{ole} E\MakeLowercase{vent} H\MakeLowercase{orizon}}

If we calculate the entropy of a thin shell of matter field
in the flat spacetime using the
method followed in this article, we find that the entropy is
proportional to the volume of the thin shell. Thus the WKB approximation
gives expected results with a thin shell of matter field. In the
Schwarzschild space this gives an appropriate flat space limit.
Thus, in this section, we will consider the entropy of a thin shell of matter field 
of a given thickness surrounding the horizon.

The solutions of the massless scalar field in terms of the Regge\textendash Wheeler
tortoise coordinates are described in [18]. A detailed discussion can be
found in [19]. The solutions are given by the following expressions:

\be
u(l,m,k|x) = {1 \over {2{\pi}{\sqrt{2k}}r}}{{R_l}(k|r)}{{Y_{lm}}(\cos \theta)}
{e^{im\phi}}{e^{-iEt}}.
\ee

\noindent{The Regge\textendash Wheeler tortoise coordinate is given by the following expression:}

\be
r^{*} = r + 2M{ln({r \over 2M} - 1)}.
\ee

\noindent{In terms of this coordinate, the radial equation takes the simple form}

\be
[{{d^2}\over{d{{r^*}^2}}} - U(l,E|r)]{{R_l}(k|r)} = 0,
\ee

\noindent{where}

\be
U(l,E|r) = -{E^2} + V(r)[{{l(l + 1) \over {r^2}}} + {{2M}\over{r^3}}].
\ee

\noindent{The potential is vanishing in the asymptotic regions corresponding to
$r \rightarrow 2M$ $(r^{*} \rightarrow {-\infty})$ and  
$r \rightarrow {\infty}$ $(r^{*} \rightarrow {\infty})$.
It follows from equations (4,13,14,15) that the WKB approximation holds
well for ${{R_l}(k|r)}$ in the near-horizon region with $k(r)$ given by
Eqn.(4) (more appropriately by Eqn.(19)). One can check this by direct substitution.
When back-scattering from the potential barrier in the intermediate
region is considered, the
near-horizon and far asymptotic solutions are given by the following
expressions:}

\ba
{\overrightarrow{R_l}}{(E|r)} \rightarrow   {e^{iE{r^*}}} + {\overrightarrow{A}(l,E)}{e^{-iE{r^*}}},  
~{{r^*} \rightarrow {-\infty}}  \\ \nonumber   
{B(l,E)}{e^{iE{r^*}}}, ~{{r^*} \rightarrow {\infty}} 
\ea

\noindent{and}

\ba
{\overleftarrow{R_l}}{(E|r)} \rightarrow   {{B}(l,E)}{e^{-iE{r^*}}} & , &  
{{r^*} \rightarrow {-\infty}}  \\ \nonumber   
{e^{-iE{r^*}}} & + & {\overleftarrow{A}(l,E)}{e^{iE{r^*}}}, ~{{r^*} \rightarrow {\infty}} 
\ea

\noindent{The arrows indicate the incoming and outgoing waves.
The constants $A$ and $B$ satisfy definite conditions given by Eqn.(105) to
Eqn.(108) in [18]. We will later give a qualitative discussions on the relative
magnitudes of these coefficients. 
The solutions of the form ${e^{iE{r^*}}}$ are vanishing at the horizon for
all positive definite $E$ and thus give a continuous spectrum. This feature
is also observed in the optical metric approach (see [17] and the references
therein). Mathematically, the brick wall cut-off parameter acts as a regularising parameter
on the continuous energy spectra associated with solutions vanishing at the horizon [18].   
The corresponding solutions are given by the sinusoidal functions of $(k{r^*})$ in place
of being purely exponential. Note that the solution ${e^{-iE{r^*}}}$ is not well-behaved
at the horizon [19].}

We consider a thin shell of matter field of a given thickness surrounding the horizon [3]
and is confined near the horizon. This gives us the dependence of the 
entropy only on the near-horizon states. We can impose the following boundary conditions
on the matter field:

\be
{\psi(2M + h)}, {\psi(2M + d + h)} = 0.
\ee

\noindent{We again assume $h << 2M$ and $d << 2M$. We also assume that the thickness
of the thin shell is large compared with the Planck length. As we will find in the
following, this makes $d >> h$. We will discuss these issues at the end of this
section [see Eqn.(33)]. We can hereafter
calculate the scalar field entropy using the semiclassical Bohr\textendash Sommerfeld
quantization rule. In this case, there is no restriction on the upper
value of the energy integral. However the infinite upper limit is not 
much significant due to the smallness of the Hawking temperature and
the associated exponential suppression. We will justify the lower
limit $(E = 0)$ later in this section. The scaling arguments given in the second
paragraph below Eqn.(11) also apply in the following calculations and 
the comments regarding the form of the dependence of the scalar field entropy on the black hole 
parameters remain the same. 
For the actual solutions, it is expected that the radial range throughout which the solutions
are stationary will decrease as we decrease the value of $n$. 
Thus, a better approach would be to consider the
modes that are stationary for any value of $r$ between the interval from $r = 2M + h$ to
$r = 2M + d + h$. In this case
the upper limit of the energy integral is expected to be restricted to a finite value as
the range of the radial integration is small. This is more appropriate
if we consider the compatibility of the boundary conditions and the back reaction
problem. However, as mentioned above, we will later find that owing to an exceedingly large thermal suppression factor, 
we can take the upper limit of the energy integral to be infinity and
the difference between these two approaches is expected to be negligible. 
We will discuss this issue later.
The continuous character of the energy spectrum for the solutions
that are vanishing at the horizon and the large magnitude of the upper limit of the allowed
values of the angular quantum number in the near-horizon region [can be
found from Eqn.(20)] give a finite expression for the radial quantum
number ($n$) and the entropy. This also allows us to replace the sum over
energy eigenstates by a corresponding integral as far as the free energy
calculation is concerned. 
For computational convenience we will consider the first approach.
In this context it will be interesting to consider wave packets confined in the
near-horizon region formed from the solutions given at the beginning of the
present section.}

To calculate the entropy, we can proceed similarly to the preceding
sections. In this case it would be more appropriate if we use the
following expression of $k(r)$ as follows from Eqn.(15):

\be
k(r) = {1\over{V(r)}}{\sqrt{{E^2} - {V(r)}[{{l(l + 1)} \over {r^2}} + {{2M} \over {r^3}}]}}.
\ee

\noindent{However, the difference between Eqn.(4) and the 
above equation is negligible for the following calculations,
and we will use Eqn.(4) for $k(r)$ for counting of the states.
We will later use the above expression to estimate the ground state energy.}

Within the region of interest, the maximum of
${{V(r)} \over {r^2}}$ occurs at $r = 2M + d + h$ and to apply the
Bohr-Sommerfeld quantization rule, the angular quantum number $l$ should be
less than the value given by the following equation:

\be
J(J + 1) = {{{E^2}{(2M + d + h)^3}} \over {d + h}} \approx
{{8{E^2}{M^3}} \over {d}}.
\ee

\noindent{If we use Eqn.(19), we will get a correction term, $-1$, which
is negligible compared with the right hand side. As before, we can consider 
the upper limit of the $l$ -integration to
be given by $sJ(J + 1)$. In the near-horizon region, $s << 1$, as follows
from the discussions below Eqn.(5). This can be seen by direct substitutions. 
With this choice, $k(r)$ is real throughout the region of interest.
In this case the free energy calculation is slightly cumbersome.
We will have finite near-horizon contributions
from the higher order terms in $'s'$. In the case of the free energy of a thin shell of scalar field of 
finite width, ${h << d << 2M}$, the leading-order term in the brick wall cut-off 
parameter comes from the linear-order term in $s$. This term is given by,}

\be
F = -[{2s}ln({d \over h})]{{{8M^4}{{\pi}^3}} \over {15 {{\beta}^4} d}} - {{s{\pi}^3} \over {30{\beta}^4}}{(2M)}^3.
\ee

\noindent{The leading order-entropy is given by the following expression:}

\be
S = {{4{\pi}^3 s} \over {15d}}(2M){({{2M} \over {\beta}})^3}ln({d \over h}).
\ee

\noindent{We will later discuss the higher order terms in $s$.
The above expression is similar to the expression of the scalar field entropy obtained
by 't Hooft [3]:}

\be
S = {{8{\pi}^3} \over {45 h}}{2M}({{2M} \over {\beta}})^3 .
\ee

\noindent{However, as we have found in the
earlier sections, the leading-order divergent term in the brick wall
cut-off parameter is again only logarithmically divergent, unlike an inverse
power divergence obtained earlier. The logarithmic divergence is expected 
if we consider the form of the solutions in the near-horizon region given by Eqs.(12) to (15),
the WKB quantization rule with $k(r)$ given by Eqn.(4) or Eqn.(19), the 
infinite redshift near the horizon and the form of the proper distance in the near-horizon region.}

Both the brick wall cut-off parameter and the thickness are radial parameters and we can
replace them by the covariant cut-off parameters ${\epsilon_d}$ and 
${\epsilon}_h$ given by the following expression:

\be
\epsilon = {{\int}_{2M}^{2M + h}}{{dr} \over {\sqrt(1 - {{2M} \over r})}}.
\ee

\noindent{For $h << 2M$, we have}

\be
h = {{{\epsilon}_h}^2 \over {8M}},
\ee

\noindent{and a similar expression for ${{\epsilon}_d}$.
The expression for the entropy is given by}

\be
S = {{32s{\pi}^3} \over {15{{\epsilon_d}^2}}}{(2M)^2}{({{2M} \over {\beta}})^3}
{ln({{{\epsilon}_d} \over {{\epsilon}_h}})}.
\ee

\noindent{If we now consider $\beta$ to be given by the inverse of the Hawking temperature,
$\beta = 8 \pi M$, we have}

\be
S = {{s} \over {30 \pi {{\epsilon_d}^2}}}{ln({{{\epsilon}_d} \over {{\epsilon}_h}})}{A \over 4}.
\ee

\noindent{Thus, we have a universal expression of the matter field entropy which is proportional
to the area of the event horizon. If we now equate the above expression to the
Bekenstein\textendash Hawking entropy [3], we find that the value of the covariant
cut-off parameter becomes independent of the mass of the black hole and is
given by the following expression:}

\be
{\epsilon_h} = {{\epsilon}_d}{e^{-{{30 \pi {{\epsilon_d}^2}} \over {s}}}}.
\ee

\noindent{This value is dependent on $s$ and the thickness of the thin-shell, $d$. 
We will discuss this issue in detail at the end of this section.}

The above expression is different from the corresponding expression obtained by 't Hooft,

\be
\epsilon = \sqrt{1 \over {90 \pi}}.
\ee

\noindent{The internal energy of the thin-shell scalar field can be obtained
from the well known expression:}

\be
U = {{\partial} \over {\partial \beta}}(\beta F),
\ee

\noindent{and with the above value of ${\epsilon_h}$ is given by the following expression:}

\be
U = {3 \over 8}M.
\ee

\noindent{It is easy to find the corresponding expressions for an asymptotically flat
nonextreme Reissner\textendash Nordstrom black hole background. In this case, the inverse
of the Hawking temperature and the covariant cut-off parameter are given by
$\beta = {{4 \pi {{r_+}^2}} \over {r_+ - r_-}}$ and
${{\epsilon}^2} = ({{4 {{r_+}^2}} \over {r_+ - r_-}})h$ respectively.
The entropy of the thin shell of scalar field surrounding the horizon
is given by the following expression:}

\be
S = {{s} \over {30 \pi {{\epsilon_d}^2}}}{ln({{{\epsilon}_d} \over {{\epsilon}_h}})}{A \over 4}. 
\ee

\noindent{The thin shell scalar field entropy is again
proportional to the area of the horizon when the temperature is given by the Hawking
temperature. This expression gives a consistent value when the $Q = 0$ Schwarzschild limit is considered.
We can do similar calculations for a thin shell of matter field in the flat space-time.
In this case the entropy becomes proportional to the volume of the thin shell in the limit when
the thickness is small compared with the radius of the inner surface.
In the black hole backgrounds the entropy
of the thin shell is proportional to the horizon area if we express the entropy
in terms of the covariant cut-off parameter.
The free energy and entropy
of the thin shell are logarithmically divergent in the 
covariant cut-off parameter.}

The covariant cut-off parameter is given by Eqn.(27) if we
equate the thin shell scalar field entropy to the Bekenstein\textendash Hawking entropy.
Thus the value of $'\epsilon_h'$ given by Eqn.(27) is independent of the
black hole parameters. This is expressed in terms of the Planck length. 
We now briefly illustrate the values of the thickness of the thin shell
and the brick wall cut-off parameter. The covariant 
cut-off parameter is given by the following expression:

\be
{\epsilon_h} = {{\epsilon}_d}{e^{-{{30 \pi {{\epsilon_d}^2}} \over {s}}}}.
\ee

\noindent{We choose the thickness of the thin shell to be large
compared with the Planck length. This allows us to use the quantum field theory
in curved spaces to find the thermodynamical properties of the matter field if we 
neglect the quantum gravitational effects associated with 
the strong curvature near the horizon. We take the value of 
${{\epsilon}_d}$ to be ${{10}^{-10}} cm$. In terms of the radial coordinates,
the corresponding thickness is small compared with the Schwarzschild radius for the macroscopic black holes. 
We can choose $s \sim {{10}^{-2}}$. 
In terms of the Planck length, the covariant cut-off parameter is
then given by the following expression:}

\be
{\epsilon_h} = [{{10}^{24}}]{\exp({-{{10}^{52}}})}.
\ee

\noindent{This is an extremely small quantity and the brick wall can be
considered to be almost coincident with the horizon. 
This is expected if we consider the
solutions of the wave equation given in this section and is also consistent
with the observed behaviours of the solutions in the optical metric
approach. In both the cases it is found that there are solutions
that are vanishing at the horizon [17,18,19]. The above
equation also justifies Eqn.(20).
The internal energy of the scalar field is less than
the mass of the black hole and is always given by Eqn.(30),{\it i.e.}
$U = {{3}\over 8}M$. This behaviour is
different from those discussed in [3] where we can not push the brick wall
arbitrarily close to the horizon due to the divergence of the scalar field internal energy. 
Similar restrictions remain valid if we follow the procedure of [3] with
a thin shell of matter field surrounding the horizon whose thickness is small
compared with the radius of the horizon but large compared with the brick wall 
cut-off parameter.}

We now consider the higher order terms in $s$ of the scalar field entropy. 
It can be seen that in the approximation $h << d << 2M$, the entropy is given 
by the following expression:

\be
S = {1 \over {30 \pi {{\epsilon_d}^2}}}[{s}{ln({{{\epsilon}_d} \over {{\epsilon}_h}})} -
{\chi(s)}]{A \over 4},
\ee 

\noindent{where ${\chi(s)} = {{{s^2} \over 4}[1 + {s \over 12} + ...]} > 0$ for $s > 0$.
For $d >> h$, the second term is negligibly small compared with the first term.
We will later find that this term may be significant in the context of the Euclidean sector
calculations of the scalar field entropy.}

In the thin shell model, we have considered the solutions that are vanishing at the two boundaries
and are stationary throughout in between. The justifications
are the behaviour of the entropy of a thin shell of scalar field obtained by a
similar method in the flat space using the spherical polar coordinates
and the form of the solutions in the near-horizon region.
The thin shell boundary condition and the boundary conditions discussed in the Section:II 
are also important if we want to compare the Lorentzian sector calculation with the
Euclidean sector calculations. We will discuss this issue later in this section.
In obtaining the expression of the free energy given by Eqn.(21), we have assumed 
that the lower limit of the energy integral in Eqn.(8) is zero. We can
estimate the lower limit by putting $n = {1 \over 2}$, ${r_1} = 2M + h$, 
$L = 2M + d + h$ and $l = 0$ in Eqn.(3) with $k(r)$ given by Eqn.(19). 
As earlier we take ${{\epsilon}_{d}} \approx 
{{10}^{- 10}} cm$. The lower limit of the energy integral then becomes of the order
of ${{10}^{-52}}{E_P}$. Here ${E_P}$ is the Planck energy. In this case
${{\beta}E} \sim {{10}^{-12}}$. Here the mass of the black hole is taken to be ten times the
solar mass. These discussions justify taking the lower limit of the energy integral in Eqn.(8) to
be zero. These approximations remain to be good with massive particles.
We can have an upper estimate of the energy by considering the total
internal energy given in Eqn.(30). This expression is independent of $d$ in the
approximation considered in this section. If we choose a value of the energy 
$\sim {M{c^2}}$ with $M$ being ten times the solar mass, the thermal suppression factor 
becomes $\sim {\exp(-{10}^{76})}$. On the other hand, for the rest energy of the 
electron, the thermal suppression factor is $\sim {\exp(-{10}^{17})}$. 
Thus, we can safely consider the upper limit of the energy integral in Eqn.(8) to be infinite. 
These comments justify the scaling
arguments given in the second paragraph below Eqn.(11) of Section:II, which justify using the same value of the 
maximum energy to assign the maximum value of $l(l + 1)$ for all the radial quantum 
numbers below $n$. Although the actual numerical coefficient will be different, the area law
of the thin-shell scalar field entropy obtained by using the above approximations is expected to hold
well. The above arguments remain valid for the massive particles whose masses
are small compared with the black hole mass. This is expected from the near-horizon geometry.
Instead 
of using the rigid wall boundary conditions,
we can consider solutions which are vanishing at the brick wall and are only stationary 
for any value of the radial variable below $r = 2M + h + d$. Where $d << 2M$ but $d$
is large compared to the Planck length. This would be similar to a particle 
in a half-infinite potential well and is more realistic
in the present context. 
The radial quantum numbers are again given by Eqs.
(3) and (19) with an appropriate upper limit. In general, for given values
of the thickness $d$ and a suitable maximum energy $E$, the maximum allowed value
of $l(l + 1)$ is determined as before. Note that we are interested to calculate the density of the states and not the exact eigenvalues.
This maximum values of $l(l + 1)$ and $E$
determines $n$ from the WKB quantization rule. The maximum energy and the
upper limit of the radial integration in the WKB quantization rule are
expected to decrease as we decrease $n$. We can evaluate $n$ at a suitable value of $d$,
multiply it by the corresponding angular degeneracy factor $(2l + 1)$ and integrate over $l$ to 
find the total number of solutions with energy less than or equal to $E$. In this
scheme, every value of $n$ (corresponding to different values of the upper limit
of the radial variable) is assigned the same angular degeneracy factor evaluated
at $r = 2M + h + d$. In the near-horizon region the angular degeneracy factor increases 
quadratically with decreasing proper distance ${{\epsilon}_{d}}$. On the other
hand the maximum value of the energy decreases with $r$ and the angular degeneracy 
factor decreases quadratically in the energy. The expression (26) of the
entropy indicates that the entropy remains to be logarithmically divergent
in the brick wall cut-off parameter provided the upper limit of the 
radial integral in the quantization rule is not of the order of the brick wall parameter itself. 
The lower value of $d$ is restricted by the
ground state energy corresponding to a ${1 \over 2}$ term in the WKB quantization rule.
In the flat space, the centrifugal potential blows up as we approach the origin if the angular momentum is finite. Thus, for a spherically symmetric potential, the ground state usually has $l = 0$. In the present case, eqn.(4) indicates that the centrifugal term in the square root
vanishes as we approach the horizon.
However, the two terms in the square root are comparable as long as the spatial extension of the ground state is not of the order of the brick wall cut-off itself.
We can set ${\epsilon_d} = {10^{-15}}cm$ for the ground state. This is a limiting distance to apply the semiclassical theory of QFT in
curved spaces. It is expected that the angular quantum number will be zero for the ground state. 
In this context, it is important to note that the WKB approximation holds good for high values of the radial quantum number. The significant contribution to the entropy is expected to arise from the high values of the radial quantum number. Thus, the scalar field entropy is expected to be logarithmically divergent in the
brick wall cut-off parameter and inverse quadratically dependent on the proper thickness of the thin shell. 
As far as the energy integral is concerned,
arguments similar to those given above justify the same values for the limits as were chosen
for the thin shell model and we can again apply the scaling arguments given in
section:II in favour of the area law obtained in this section. 
One should appreciate the significance of the near-horizon geometry for
the arguments given in this paragraph. We also note
that the solutions which are stationary upto ${R_1} = d$ from the brick wall 
will again become stationary at ${R_2} \approx {{8{M^2}} \over {\epsilon_d}}$ for
${\epsilon_d} << 2M$ which is the case in the present section. 
This is determined from the relation: 
${{V(R_1)} \over {{R_1}^2}} = {{V(R_2)} \over {{R_2}^2}}$. This is a large
value for the macroscopic black holes and it is extremely less probable for these 
solutions to tunnel through the intermediate potential barrier to infinity. These 
facts and the area law obtained in this section make the study of the states stationary 
in the near-horizon region particularly significant.    
All the above discussions also indicate that the thin shell model is a good approximation 
to find out the expression of the entropy of a scalar field confined near the horizon. 
Lastly we can consider the artificial
case of a thin shell of width of the order of the brick wall cut-off parameter
itself [32]. It is obvious from Eqn.(26) that the scalar field entropy is then quadratically 
divergent in the proper distance thickness, ${{\epsilon}_{d}} = ({\sqrt{2}} - 1){{\epsilon}_{h}}$, 
in the limit ${{\epsilon}_{d}} \rightarrow 0$. However, the proper thickness becomes 
of the order of the Planck length when we equate the scalar field entropy 
to the black hole entropy and it is not appropriate to use the semiclassical 
approximation of quantum fields in curved spaces in this case.

The black hole entropy may be interpreted as originating from a
cluster of matter field confined in the near-horizon region [3].
This is important in the context of
explaining the black hole entropy in terms of the near-horizon states [4,18,24,25,26,27] and
also for the entanglement entropy approach to explain the black hole entropy [11,23,33].
The most significant aspect of the calculations given in this section is that the
entropy of a thin shell of scalar field surrounding the horizon is proportional
to the area of the horizon and is logarithmically divergent in the brick wall cut-off parameter. 
The situation for the extreme Reissner\textendash Nordstrom black hole is not much illuminating
and the scalar field entropy vanishes when the Hawking temperature is taken to be zero.
The present article indicates that the fixed point set character
of the event horizon and the form of the metric are significant for the nontrivial features of
the entropy of a thin shell of scalar field surrounding the horizon. We should
note that the term 'thin' refers to a proper thickness which is of the order
of the atomic scale and small compared with the Schwarzschild radius of the macroscopic
black holes. The leading-order entropy is proportional to the horizon surface
area and not related to the proper volume of the thin shell. It is obvious
from Eqn.(21) that the free energy vanishes in the flat space limit $M \rightarrow 0$.
This is expected since the corresponding region does not exist in the flat space
limit. We note in passing that
one should be careful in considering the effect of the redshift in interpreting the
expression of the entropy of a thin shell of scalar field even when the inner radius is at the
other side of the potential barrier. We also note that the WKB approximation
holds well when $V(r)$ is small.

To the leading-order terms, the expressions of the scalar field entropy 
obtained in this article is different from those obtained by the other approaches mentioned in the introduction. 
The different approaches to calculating
the matter field entropy in a curved space-time use different regularization schemes
and have different mathematical approximations.
The brick wall model approach is based on the Lorentzian sector of the black hole
space time and is expected to give the most robust expression. 
It is convenient to use the Zeta function regularization
scheme in the Euclidean sector calculation for the partition function
of a scalar field in a black hole background. This is considered in [27,34,35] with the
boundary conditions similar to those used in this article.
In the Zeta function regularization scheme to evaluate the partition function, 
the partition function is given by the Eqn(3.2) of [27]:

\be
\ln[Z] = {1 \over 2}[{\zeta}'(0) + \ln({1 \over 4}{\pi}{{\mu}^2}){\zeta}(0)].
\ee

\noindent{Here $\mu$ is a normalization factor and $\zeta(s)$ is the Zeta function.
Unlike the Lorentzian flat space-time, it is evident from the Eqs.(3.3) and (3.4) of [27] that in
the Euclidean sector, the momentum space integral is unbounded for each allowed value
of the corresponding energy.
In four dimensions the zeta function converges for $Re(s) > 2$. However, it
can be analytically extended to a meromorphic function with poles at $s = 1, 2$ 
and is regular at $s = 0$. Thus, to extract a well-defined value for the partition function (apart from an infrared divergent
part), we first evaluate $\zeta(s)$ for $Re(s) > 2$ and thereafter analytically continue the 
resulting expression to $s = 0$.}

In a curved spacetime $\zeta(s)$ is given by the following
expression:

\be
\zeta(s) = {1 \over {\Gamma(s)}}[{{\int}^{1}_{0}}{t^{s - 1}}Y(t)dt + {{\int}^{\infty}_{1}}{t^{s - 1}}Y(t)dt],   
\ee

\noindent{where $Y(t)$ is related to the five dimensional heat kernel [27] and $t$ is the time
in a five-dimensional manifold. The above equation
represents a Mellin transform of $Y(t)$. The evaluation of the heat kernel depends on the
WKB approximation [18]. The second integral converges and the first integral is given 
by the following expression:}

\be
{\zeta(s)}_{1} = {1 \over {\Gamma(s)}}[\sum {{{B}_{n} + {C}_{n}} \over {n + s - 2}}],
\ee

\noindent{where the coefficients $B_{n}$ and ${C}_{n}$ are integrals of the scalar polynomials
of the metric and the induced metric over the manifold and a boundary in the manifold
respectively. The details are given in [27] and [18]. 
The boundary conditions imposed on the field are similar to those chosen in
this article. One can obtain an expression for 
the inverse of the Gamma function by using the Weierstrass's formula.
For the Schwarzschild black hole,
the scalar field entropy does not contain a term proportional to the area of the horizon.
However, we can introduce conical singularity in the Euclidean manifold by identifying the imaginary
time with a period different from that given by the Hawking temperature [7,17,28]. 
The evaluation of the constants ${B}_{n}$ in presence of the conical singularity
is described in [17]. In this
case the scalar field entropy contains a term proportional to the area of the horizon but
is negative: $- {{A_h} \over {48 \pi}}$. 
This is similar in form only with the subleading term  
of the Lorentzian sector expression (34).
A different approach would be to calculate the integral (36) directly.
The part of the entropy proportional to the area is then given by 
${{A_h} \over {48 \pi} {t_0}}$, where ${t_0} \rightarrow 0$.
In this case, the scalar field entropy also contains a term which is
quadratically divergent in ${t_0}$ as ${t_0} \rightarrow 0$. This term is not proportional
to the area of the horizon. A few similar Euclidean sector calculations
are discussed in the references [7,17,25,26]. The scalar field entropy
is found out to be proportional to the horizon surface area and first order
divergent in the proper time of the five dimensional manifold used to evaluate
the heat kernel. As mentioned above, there are other terms that are more divergent and
not proportional to the horizon area. The first order divergence is then changed to a
second order divergence so as to agree with the Lorentzian sector expression
obtained by 't Hooft [3]. One can also change the first order divergence to a
zeroth order divergence so that the expression of the scalar field entropy   
agrees with the expression obtained in the present article.
The differences between the Lorentzian sector calculation and
the Euclidean sector calculations is due to the differences in the spectrum. 
One should also consider the differences between the two
geometries. The mathematical approximations used in the Euclidean sector regularization scheme
are not always well-defined [27]. They
are justifiable as long as they reproduce the Lorentzian sector expressions for
the corresponding thermal quantities.
These discussions call for a possible improvement of the Euclidean sector
regularization schemes used to evaluate the thermodynamical properties of matter
fields in curved spaces. We should also note that the continuation
to the Euclidean sector is not always trivial [36].  
We will discuss these issues later.}

A related field is to determine the expressions for the entanglement entropy
of a scalar field in a given spacetime. The 
entanglement entropy of the matter fields are found from the reduced density
matrices obtained by summing over the degrees of freedom confined within 
a given spatial region [11,23] of the flat space time. This is not exactly the same as the entropy
of the matter fields in the black hole backgrounds.   
The scalar field entanglement entropy contains a part 
proportional to the area of the boundary of the given region and is quadratically 
divergent in a cut-off parameter. An extension of this approach in the black hole
backgrounds is discussed in [33]. In this context, it will be
interesting to consider the entanglement entropy of a thin
shell of scalar field surrounding the horizon and compare
the result with the expression obtained in this article. 
This may be important to explain the black hole entropy in
terms of the near horizon physics. As is mentioned
earlier, it is extremely less-probabale that the modes 
considerd in this section will tannel through the intermidiate
potential barrier to infinity.

The scalar field entropy have been interpreted by some authors as a quantum correction
to the black hole entropy [4,9,17,24] and thus giving infinite renormalization
to the gravitational constant $G_{B}$. The one-loop effective action of a
scalar field in a curved space-time can be found by using the DeWitt\textendash Schwinger proper time representation.
The divergent parts are given by equation (6.44) in [24]. These terms are divergent only when
the space-time dimension is four and may be interpreted as giving infinite renormalizations   
to the different coupling constants present in the Einstein-Hilbert action for the gravitational
field itself. The divergent term is of the form $1 \over {(n - 4)}$, where $n$ is the space-time
dimension. The renormalized coupling constants are given by equation (6.49) of [24]. 
The divergent part of the one-loop effective scalar field lagrangian is related to the short-distance
high energy modes. The near-horizon divergence of the scalar field entropy in 
the black hole backgrounds is related to the divergence of the density of the states
associated with the fixed point character of the horizon and infinite red-shift.
In the path integral approach [27], these two calculations are related by the
continuation from the Lorentzian sector to the Euclidean section. 
The differences between the expressions of the scalar field entropy obtained
by these two different approaches, as mentioned in the
preceding paragraph, may indicate that the continuation from the Lorentzian sector to the Euclidean section
is not trivial. 
We can consider the scalar field entropy as giving quantum corrections to the black
hole entropy and thus renormalizing the gravitational constants. However, the renormalizations
to the gravitational constants obtained in this way may differ from the corresponding
expressions obtained from the one-loop effective action of the scalar field as mentioned
earlier. If we interpret the scalar field entropy in this way, 
the renormalization to $G_{B}$ obtained by t'Hooft is quadratically
divergent in the covariant brick wall cut-off parameter while the corresponding expression
obtained in this article with an improved counting of the states is only logarithmically
divergent. These are different from the $1 \over {(n - 4)}$ renormalization to $G_{B}$
coming from the one-loop effective action of the scalar field.
A treatment where both the approaches may give the same renormalizations
to $G_{B}$ had been considered in [37] using the Pauli\textendash Villers regularization scheme.
However, the one-loop effective action is quartically divergent in this approach 
as compared with the linear divergence as mentioned earlier. The present article is
relevant in this context.

\section {C\MakeLowercase{onclusions}}

To summarize, in this article we have obtained a new expression of 
the entropy of a minimally coupled scalar field in the Schwarzschild
black hole background. We have
used the brick wall model of 't Hooft. The density of states
is calculated using the WKB approximation. We found that the entropy
associated with the solutions that are stationary throughout the region of interest
is logarithmically divergent in the brick wall cut-off
parameter and is not proportional to the area of the horizon when the temperature
is given by the Hawking temperature. The divergence is associated with the intersection of the
constant-time foliations at the horizon which is a two-dimensional fixed
point set of the timelike Killing vector. We thus considered the entropy
of a thin shell of scalar field with a given thickness surrounding the black hole horizon. 
The thickness is chosen to be of the order of the atomic scale.
We have obtained an expression for the matter field entropy which is
proportional to the horizon surface area. This is valid for both the Schwarzschild and the
non-extreme Reissner\textendash Nordstrom black holes and the proportionality
constants are the same. The leading-order term in the cut-off 
parameter is again logarithmically divergent. 
We also found that the internal energy of the scalar field is less than the mass of the black hole
even if the brick wall is pushed almost onto the black hole horizon.
These aspects are different from
the earlier results where the divergent term of the scalar field entropy associated 
with the proper distance brick wall cut-off has a second order divergence
and the brick wall cannot be put very close to the horizon 
due to the back reaction problem. 
In the thin shell model we have considered the modes that are vanishing at the 
two boundaries and stationary throughout in between. We can
generalize these calculations to find the entropy of a scalar field confined
in the near-horizon region and stationary for any value of 
the radial variable from the brick wall up to a proper
distance which is again of the order of the atomic length.   
The leading order term in the cut-off parameter is again logarithmically
divergent and proportional to 
the area of the horizon.

The entropy calculations involve some approximations as 
discussed below Eqs.(11) and (34). In particular, the expressions
of the entropy can be taken to give us the correct forms and not the
exact numerical values. We have also discussed the appropriateness 
of the thin shell model to evaluate the entropy of a scalar field confined 
in the near-horizon region. We also found that it is extremely less probable 
that the solutions stationary in the near-horizon region of 
proper width of the order of the atomic scale will tunnel
through the intermediate potential barrier to infinity. This aspect together
with the area law obtained in section:III make these modes important
in explaining the black hole entropy from the near-horizon geometry.
The comparison between different approaches to evaluating the scalar field 
entropy indicates that the Euclidean sector calculations are not in agreement with the
Lorentzian sector expression obtained for the scalar field entropy. This
remains to be the situation also with the Lorentzian sector expressions
obtained by 't Hooft [3]. Thus, the Euclidean sector regularization schemes 
may need to be improved. The discussions in the Section:III are for the Schwarzschild 
and non-extreme Reissner\textendash Nordstrom black holes.
The present article indicates that the fixed point set character
of the event horizon and the form of the metric are significant for 
non-trivial features of the scalar field entropy. 
Thus, it will be interesting to use the methods described 
in this article to calculate the scalar field entropy
in the Taub\textendash NUT space and the BTZ black hole [9,10] containing nontrivial fixed
points. The WKB approximation will be particularly useful for these backgrounds. 
The discussions in this article are also significant
for a spce-time with cosmological event horizon [38].
The results of this article may be relevant 
to the concept of relative entropy [39].

\section{A\MakeLowercase{cknowledgements}}

I am thankful to the reviewer and the 
staff of 'Journal of the Physical Society
of Japan' for a few improvements.

\newpage

\section*{R\MakeLowercase{eferences}}

[1] J. Bekenstein, Phys. Rev. D {\bf 7}, 2333 (1973).

[2] S. W. Hawking, Commun. Math. Phys. {\bf 43}, 199 (1975).

[3] G. 't Hooft, Nucl. Phys. B {\bf 256}, 727 (1985).

[4] L. Susskind and J. Uglum, Phys. Rev. D {\bf 50}, 2700 (1994).

[5] A. Ghosh and P. Mitra, Phys. Lett. B {\bf 357}, 295 (1995).

[6] I. Ichinose and Y. Satoh, Nucl. Phys. B {\bf 447}, 34 (1995).

[7] S. N. Solodukhin, Living Rev. Relativity {\bf 14}, 8 (2011).

[8] R. G. Cai and Y. Z. Zhang, Mod. Phys. Lett. A {\bf 11}, (1996).

[9] K. Ghosh, Phys. Rev. D {\bf 67}, 124027 (2007).

[10] B. Mukhopadhyay and K. Ghosh, Class. Quant. Grav. {\bf 25}, 065006 (2008).

[11] L. Bombelli, R. K. Kaul, J. Lee and R. Sorkin, Phys. Rev. D {\bf 34}, 373 (1986).

[12] S. P. de Alwis and N. Ohta, Phys. Rev. D {\bf 52}, 3529 (1995).

[13] M. R. Setare, Int. J. Mod. Phys. A {\bf 21}, 6183 (2006).

[14] K. Ghosh, Nucl. Phys. B {\bf 814}, 212 (2009). 

[15] G. W. Gibbons and S. W. Hawking, Commun. Math. Phys. {\bf 66}, 291 (1979).

[16] S. Carlip, Phys. Rev. D {\bf 51}, 632 (1995).

[17] V. Frolov and D. Furasarv, Class. Quant. Grav. {\bf 15}, 2041 (1998); D. Fursaev and S. Solodukhin, 
     Phys. Rev. D {\bf 52}, 2133 (1995).

[18] B. S. DeWitt, Phys. Reports {\bf 19}, No.6 (1975).

[19] D. G. Boulware, Phys. Rev. D {\bf 11}, 1404 (1975).

[20] D. Birmingham, K. S. Gupta and S. Sen, Phys. Lett. B {\bf 505}, 191 (2001).

[21] J. Engle, K. Noui and A. Perez, Phys. Rev. Lett.{\bf 105}, 031302 (2010). 

[22] A. Corichi, J. Diaz-Polo and E. Fernandez-Borja, Class. Quant. Grav. {\bf 24}, 243 (2007).

[23] M. Srednicki, Phys. Rev. Let. {\bf 71}, 666 (1993).

[24] N. D. Birrel and P. C. W. Davies, Quantum Fields in Curved Space (Cambridge University Press, Cambridge, 1982)

[25] G. Cognola, L. Vanzo and S. Zerbini, Class.Quant.Grav. {\bf 12}, 1927 (1995).

[26] G. Cognola, Phys. Rev. D {\bf 57}, 6292 (1998).

[27] S. W. Hawking, Commun. Math. Phys. {\bf 56}, 133 (1977).

[28] S. W. Hawking, G. T. Horowitz and S. F. Ross, Phys. Rev. D {\bf 51}, 4302 (1995).

[29] R. C. Tolman, The principles of Statistical Mechanics, (Dover Publications, New York, 1979).

[30] J. J. Sakurai, Modern Quantum Mechanics (Addison-Wesley Publishing Company, Massachusetts, 1994).

[31] K. Ghosh, arXiv:0902.1601 (2014). 

[32] K. Ghosh, J. Phys. Conf. Ser. {\bf 410}, 012137 (2013).

[33] C. Callan and F. Wilczek, Phys. Lett. B {\bf 333}, 55 (1994).

[34] D. B. Ray and I. M. Singer, Adv. in Math. {\bf 7}, 145 (1971).

[35] P. B. Gilkey, Adv. in Math. {\bf 15}, 334 (1975).

[36] J. Zuo and Y. Gui, Int. J. of Theor. Phys. {\bf 38}, No.2 (1999);
     
     P. Cartier and C. DeWitt-Morette, Functional Integration: Action and Symmetries
     
     (Cambridge Monographs on Mathematical Physics, Cambridge, 2010); R. J. Rivers, Functional Integral 
     
     Methods In Quantum Field Theory (Cambridge University Press, Cambridge, 1987).

[37] J. G. Demers, R. Lafrance and R. C. Myers, Phys. Rev. D {\bf 52}, 2245 (1995). 

[38] S. W. Hawking and G. Gibbons, Phys. Rev. D 
{\bf 15}, 2738 (1977). 

[39] H. Qian, Phys. Rev. E {\bf 63}, 042103 (2001).

\end{document}